\documentclass[conference,a4paper]{IEEEtran}
\IEEEoverridecommandlockouts
\usepackage{cite}
\usepackage{amsmath,amssymb,amsfonts}

\usepackage{algorithmic}
\usepackage{graphicx}
\usepackage{textcomp}
\usepackage{xcolor}
\usepackage{subfigure}
\usepackage{multirow}
\usepackage{tabularx}
\usepackage{gensymb}
\usepackage[ruled,vlined]{algorithm2e}
\def\BibTeX{{\rm B\kern-.05em{\sc i\kern-.025em b}\kern-.08em
    T\kern-.1667em\lower.7ex\hbox{E}\kern-.125emX}}

\newcommand{\mycomment}[1]{}

\begin{document}

\title{

On the Selection of Intermediate Length Representative Periods for Capacity Expansion\\
}

\author{\IEEEauthorblockN{Osten Anderson, Nanpeng Yu}
\IEEEauthorblockA{\textit{Department of Electrical and Computer Engineering} \\
\textit{University of California, Riverside}\\
Riverside, CA \\}
\and
    \IEEEauthorblockN{Konstantinos Oikonomou, Di Wu}
\IEEEauthorblockA{\textit{System Evaluation Group} \\
\textit{Pacific Northwest National Laboratory}\\
Richland, WA \\}
}


\maketitle
\IEEEpubidadjcol

\begin{abstract}
As the decarbonization of power systems accelerates, there has been increasing interest in capacity expansion models for their role in guiding this transition. Representative period selection is an important component of capacity expansion modeling, enabling computational tractability of optimization while ensuring fidelity between the representative periods and the full year. However, little attention has been devoted to selecting representative periods longer than a single day. This prevents the capacity expansion model from directly simulating interday energy sharing, which is of key importance as energy generation becomes more variable and storage more important. To this end, we propose a novel method for selecting representative periods of any length. The method is validated using a capacity expansion model and production cost model based on California's decarbonization goals. We demonstrate that the representative period length has a substantial impact in the results of the capacity expansion investment plan. 
\end{abstract}


\section{Introduction}
\label{intro}

\subsection{Background \& Problem Statement}
Decarbonization of power grids has been identified as a critical component of the response to climate change, and many governments have adopted laws to this end. For example, the California State Legislature passed SB100 in 2018 and set the target of 60\% by 2030 and 100\% by 2045 of retail electricity sales from renewable sources \cite{SB100}. This requires substantial investment in green technology, especially in renewable generation and energy storage, with the value of investment and operation over this timeline on the order of 100s of billions of US dollars. As a result, effective planning of the investment rollout is of critical importance for minimizing cost, ensuring reliable operation, and meeting policy regulations. 

Models which enable this planning are known as capacity expansion models (CEMs), and are related to the task of generation expansion planning. These models have been used for decades, but have been subject to additional attention in recent years for their role in guiding the transition to lower-carbon, high-renewable grids. 

CEMs seek to optimize generation and transmission investment strategies, and typically model two timescales: annual investment decisions and hourly operation decisions. As a result of the hourly timescale, CEMs can quickly become intractable if all 8760 hours in a year are modeled. To address this, CEMs will typically utilize representative periods instead of all 8760 hours. The main goal in selecting these representative periods is to maximize the similarity between the annual behavior and the surrogate representation, while achieving a sufficient reduction in the associated computational load.

\subsection{Related Work \& Paper Contribution} \label{relatedwork}
The problem of representative period selection in CEMs has received much attention in the literature. The majority of the work on this topic has been built around the framework of time series clustering.
The authors in \cite{Teichgraeber} compare a variety of clustering methods. The authors in \cite{Pfenninger} present a comparison of clustering and downsampling approaches. 
In \cite{Tso}, a clustering method is proposed which requires that each cluster consist of a contiguous set of days, while a downsampling-based approach is proposed in \cite{Liu}. These clustering-based approaches all share the common drawback that they are only suitable for the selection of representative days, and not periods of multiple days in length, as will be discussed in detail later. 

In addition, several works have focused on modeling the full year contiguously to allow for the tracking of interday and long-term energy storage by reducing the modeling frequency \cite{Tejada, Pineda}. 
The key drawback to these works is that they lose the sequentiality of hours, and thus cannot model important inter-hour details, such as ramping in power plants. 

The references cited above primarily concentrate on the selection of representative days and offer general assertions regarding the algorithm's ability to choose periods of varying lengths, such as a representative day or week.
However, let us consider the case where the desired period is of an intermediate length, such as 3 days. 
The time series clustering framework upon which the majority of representative period selection algorithms are built, requires the full time series to be divided into subsequences. Clustering these subsequences becomes highly dependent on the starting point. In particular, the load exhibits significant differences in both shape and magnitude between weekdays and weekends. For example, the Euclidean distance of a Friday-Saturday-Sunday subsequence to a Saturday-Sunday-Monday subsequence would likely be large because the loads of Friday would be compared to Saturday and the loads of Sunday would be compared to Monday. On the other hand, clustering of overlapping subsequences, obtained by sliding a window across the full time series with a stride shorter than the subsequence length, has been established to return essentially random results \cite{meaningless}. 
Even in the case of representative weeks, there are considerable drawbacks. Generally speaking, capacity expansion problems tend to reduce the annual temporal coverage to roughly 10\% or less. In the case of days, this permits 37 days, but in the case of weeks, this permits only 5 weeks. Intuitively, one would expect that representing a full year is more difficult given 5 choices than 37 choices. 

Considering the limitations of existing clustering methods, along with the need to capture interday variability in CEMs subject to high levels of energy storage and renewables penetration, the selection of representative periods longer than one day becomes imperative. 
Indeed, enabling interday sharing of energy through storage modeling is a crucial, yet often ignored, aspect that would become particularly valuable during days of low generation from renewable resources.

While the selection of a representative period length is fundamentally an experimental design decision, there exists a noticeable gap in research when it comes to effectively choosing a period longer than a day but shorter than a week. To bridge this knowledge gap, we introduce a novel snippet algorithm specifically designed for selecting representative periods that extend beyond a single day. 
By comparing subsequences instead of full sequences, the proposed snippet algorithm is able to select representative periods of arbitrary length from complex datasets. 
The proposed algorithm draws significant inspiration from \cite{snippets}; however, we have made several tailored adjustments to accommodate the unique domain to which our proposed algorithm is applied. 


The remainder of this paper is structured as follows. Section \ref{technicalmethod} provides an overview of the original time series snippets algorithm, its differences from the proposed algorithm, and the proposed algorithm itself. Section \ref{experimentalvalidation} discusses the numerical study setup and results. Section \ref{conclusion} presents the conclusions.

\section{Technical Method} \label{technicalmethod}

\subsection{Overview of time series snippets} 
\begin{figure}[h]
\centering
\includegraphics[width = \linewidth]{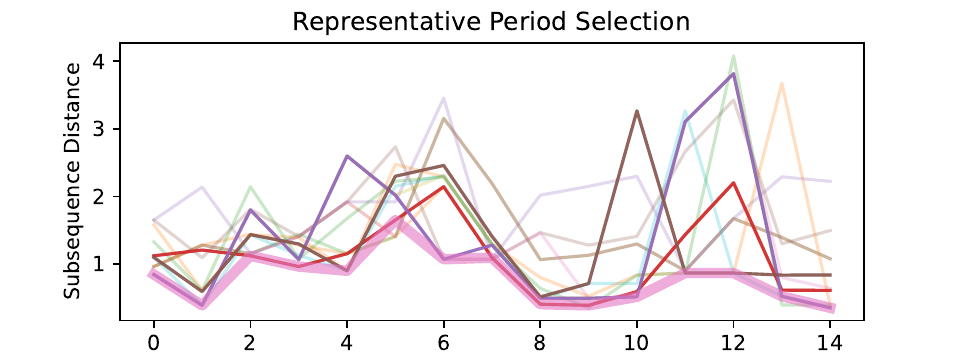}
\caption{Visualization of distance profiles and minimization of area under the curve of selected profiles.}
\label{fig:distance_vis}
\end{figure}

The discussion of time series snippets below provides a concise overview of the algorithm that served as the inspiration for the proposed method in Section \ref{proposedmethod}.
The time series snippets algorithm is built on top of the matrix profile distance (MPdist) \cite{mpdist} measure, which is in turn built on top of the matrix profile \cite{mp1}. MPdist compares two-time series and considers them to be similar if they have similar subsequences. At its most basic level, the distance is the j-th smallest Euclidean distance between subsequences. 
More specifically, the goal of time series snippets is to select, from a time series $T$ with length $t$, subsequences of length $s$ that best generalize the full-time series. First, the full-time series is separated into non-overlapping subsequences $S_i$ with $i \in [0, t/s-1]$. Each of these subsequences then has an MPdist profile $MPdist_i$ compared to the full-time series. If each $MPdist_i$ were plotted, the goal would be to select the $k$ profiles that minimize the area under the curve of the combined profiles, as shown in Fig. \ref{fig:distance_vis}. To select these representative subsequences, a greedy algorithm is proposed, choosing the subsequence that gives the greatest reduction in the cumulative distance in each iteration. 

The key contribution of these matrix profile-related methods is that they scale well to extremely long time series. The problem of representative period selection for power system planning typically considers one year of data at hourly frequency, for a time series of length 8760, which is extremely short in that context. Further, we don't need to calculate the distance measure for every subsequence, as we can exploit the known daily periodicity of our time series. For this reason, we can calculate a distance measure similar to the $MPdist$ without relying on algorithms related to the matrix profile. This also enables us to utilize overlapping subsequences as $S_i$. Finally, the problem size allows us to select the representative snippets through convex optimization rather than relying on a greedy algorithm. 

\subsection{Proposed Method} \label{proposedmethod}
\begin{figure}[h]
\centering
\includegraphics[width = 0.92 \linewidth]{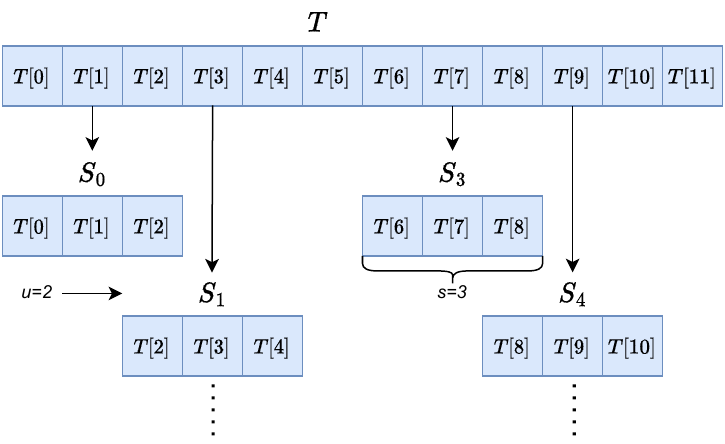}
\caption{Visualization of an example time series $T$ and subsequences $S_j$ with $u=2$, $s=3$.}
\label{fig:subseq_vis}
\end{figure}

Let $T = \{T[0], \dots T[h],\dots, T[t-1] \}$ represent the yearly multivariate time series of length $t$ and $T[h]$ be the tuple of measurements at hour $h \in [0, t-1]$. 
This tuple typically encompasses load, solar generation, and wind generation information; however, the proposed method remains agnostic to the input features, providing adaptability in the analysis. 
Let also $S = \{S_0, \dots S_j, \dots S_{m-1}\}$ be the set of subsequences, and $u$ the stride of the window that generates subsequences of length $s$. The subsequence $S_j$ is then defined as: 
\begin{equation}
    S_j = T[j\cdot u:j\cdot u+s]
\end{equation}
There will be $m=\frac{t-s}{u} + 1$ total subsequences, thus $j\in[0,m-1]$. 
These will be the candidate subsequences used for selecting representative periods. 
Similarly, we can define non-overlapping subsequences of $T$, which we will refer to as segments. There will be $n=\frac{t}{u}$ such segments, thus $i\in [0,n-1]$. The segment $T_i$ is then defined as:
\begin{equation}
    T_i = T[i \cdot u: (i+1)\cdot u]
\end{equation}
A visual representation of the definitions can be seen in Fig. \ref{fig:subseq_vis}. For clarity, $i$ will be reserved to index the time series segments $T_i$ and $j$ to index the subsequences $S_j$. 

Inspired by the MPdist, let $\mathcal{D}$ be a matrix of distances with shape $n \times m$. For our case, we will assume $u$ and $s$ are chosen such that $n$, $m$, and $s/u$ are integers. We also have domain knowledge of periodicity. Each of the features, especially load and solar generation, have strong 24-hour cycles. It is unlikely that an afternoon subsequence from one day would be similar to a nighttime subsequence from another day. Further, this is physically meaningless in the context of capacity expansion. For this reason, whereas the MPdist compares subsequences for every timestep, we apply stride $u=24$ in calculating the distances. In essence, this compares each day in the subsequence to each day in the full time series, and assigns a distance correspondingly.  
\begin{equation} \label{distance}
\begin{split}
        \mathcal{D}_{i,j} = \min_x ||S_j[x\cdot u : (x+1)\cdot u] - T_i||, x \in \left[0, \frac{s}{u}\right)
\end{split}
\end{equation}

The goal is then to find a subset of those candidate days which best captures the patterns for the year as a whole. This goal is the same as the one visualized in Fig. \ref{fig:distance_vis}.
Time series snippets were originally proposed with a greedy algorithm that iteratively selects the subsequence which minimizes the cumulative sum of distances to the full time series, necessitated by the long time series the algorithm was designed for.
Because our time series is rather short, we can instead formulate this problem as a mixed integer linear program and find the solution using any suitable optimization solver.
\begin{flalign}
    \min_{\alpha} & \sum_{i=0}^{n-1} dist_i     \label{eq:algo}
\\
     s.t. & \sum_{j=0}^{m-1} \alpha_j = k \nonumber \\
    & dist_i = \sum_{j=0}^{m-1} md_{i,j} \cdot\ \mathcal{D}_{i,j}, \; \forall i \in [0, n) \nonumber \\
    & \sum_{j=0}^{m-1} md_{i, j} = 1
        , \; \forall i \in [0, n) \nonumber \\
    & md_{i, j} \leq \alpha_j
        , \; \forall i \in [0, n), \; \forall j \in [0, m) \nonumber \\
    & dist \in \mathbb{R}^{n}, \; \; md \in [0, 1]^{n \times m}, \;\; \alpha \in [0, 1]^m, \nonumber 
\end{flalign}
where k is the desired number of representative periods; $\alpha_j$ is a binary indicator selecting subsequence $S_j$
as a representative period; $dist_i$ is the minimum distance between the selected representative periods to $T_i$; and $md_{i,j}$ is a binary indicator signifying that subsequence $S_j$ has the smallest distance to day $T_i$.
Within CEM, representative periods are typically weighted by the amount of the year that they account for. 
The weights associated with each representative period are a function of $md_{i, j}$, and can be written then as:
\begin{equation}
    w_j = \sum_{i=0}^{n-1} md_{i,j} / (s/u),
\end{equation}
where $s/u$ ensures the weights sum to 365.


\section{Experimental Validation} \label{experimentalvalidation}
\subsection{Experimental Setup}

To the author's knowledge, no paper has made a dedicated attempt to address the sampling of intermediate-length representative periods in capacity expansion planning. This absence poses a challenge when comparing the proposed method with widely used state-of-the-art approaches.
We will compare the performance of the proposed algorithm to a popular method for selecting representative days, and show that our algorithm is at least comparable with the state-of-the-art for this task.
The proposed method will also be used to compare single-day planning to multi-day planning. The goal of this comparison is to show the value in simulating representative periods longer than one day, particularly in sizing energy storage. Our goal is not necessarily to show the optimality of a particular representative period length, but rather to demonstrate the differences between period lengths on investment plans and operational cost. 

The following general experimental design will be used to validate the proposed method. First, the representative days are selected and used within the CEM. Then, the investment decisions are fixed, and the model is solved again as a production cost model.
The production cost model (PCM) is ran for the full year in two-week stages, and results from this will be referred to as fullspace results. The choice of two weeks is somewhat arbitrary, with the key being that this length is considerably longer than each of the candidate representative period lengths to avoid giving bias towards any particular length. 

\mycomment{\begin{figure}[h]
\centering
\includegraphics[width = \linewidth]{numericalStudyFlow (2).png}
\caption{Flow of experimental validation}
\label{fig:numstudy}
\end{figure}}

\subsection{Simulation Models}
The PCM and CEM used are both zonal models of the Western Interconnection, primarily focused on California. Both utilize a MILP adaptation of the formulation and data of E3's RESOLVE decarbonization model \cite{RESOLVE}. Further discussion of the decarbonization model including full formulations of the objective and constraints are available in \cite{anderson2023optimize}.
The principle goal of both models is minimizing the cost of serving load. The PCM focuses on operational decisions, including the scheduling of power plants, to minimize operating costs while satisfying operating and reliability constraints \eqref{eq:PCM}. 
Operating constraints maintain safe resource limits (e.g., power plant capacities), while reliability constraints guarantee zonal power balance and ancillary services for secure power supply.
\begin{equation} \label{eq:PCM}
\begin{aligned}
\min \quad & C_{gen}\\
\textrm{s.t.} \quad & \text{Operating constraints} \\
  & \text{Reliability constraints}
\end{aligned}
\end{equation}

The CEM essentially adds an additional level to the PCM by allowing investment in additional generation capacity. It solves operational and investment decisions to minimize the total cost of generation $C_{gen}$, maintenance $C_{maint}$ and investment $C_{inv}$ \eqref{eq:CEM}. Investment decisions include both the addition of new resource capacity and retirement of existing gas generators. Policy constraints are integral to a decarbonization model, and include emissions limits and renewable portfolio standards.

\begin{equation} \label{eq:CEM}
\begin{aligned}
\min \quad & C_{inv} + C_{maint} + C_{gen}\\
\textrm{s.t.} \quad & \text{Policy constraints}\\
  & \text{Operation constraints} \\
  & \text{Reliability constraints}
\end{aligned}
\end{equation}

\subsection{Results and Discussions}

First, we compare our method to a popular approach for representative day selection: k-means clustering using load, wind, and solar profiles with medoid cluster center representation used in the CEM. We use our proposed algorithm with subsequence length $s=24$ to select 21 representative days. Our algorithm has a total cost (investment, maintenance, fullspace operations) of 14.560 billion US dollars as compared to 14.611 billion US dollars. \cite{Teichgraeber} compared many clustering techniques for power system planning, and established that there are not clear patterns on which technique is best, and many have comparable performance. With this in mind, we can suggest that even for representative day selection, our proposed method is at least comparable with one of the most commonly used representative day selection approaches. 

With the validity of the proposed method established, we now seek to defend the motivation behind selection of longer representative periods. 
This will be explored via investment, fullspace operation cost, emissions, and the investment portfolio, considering representative periods of 1 to 5 days, i.e. with subsequence length $s \in [24, 48, 72, 96, 120]$. In each case, periods are adjusted to model a total of 35 or 36 days, nearly 10\% of the annual days.

\begin{figure}[h]
\centering
\includegraphics[width = 0.95\linewidth]{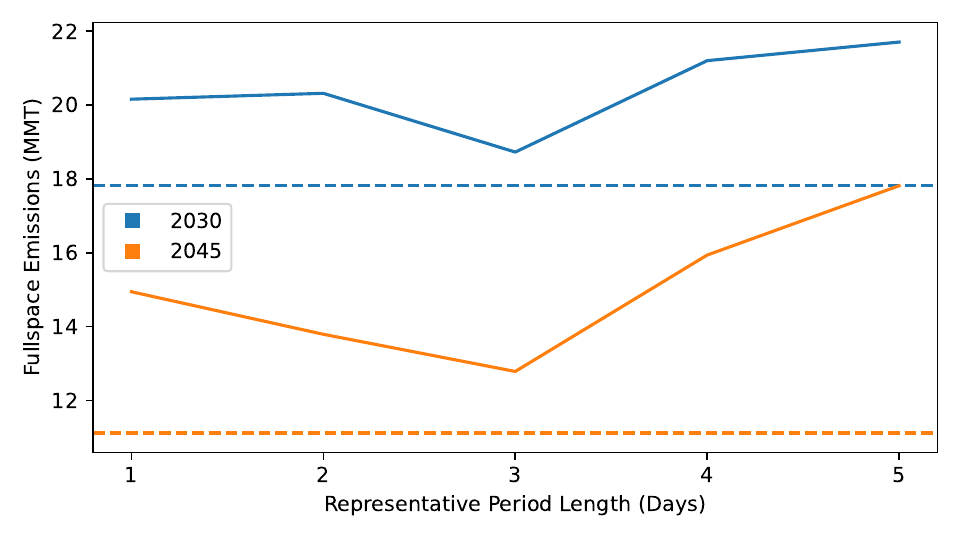}
\caption{Fullspace emissions}
\label{fig:emissions}

\end{figure}

Fullspace emissions are shown in Fig. \ref{fig:emissions}. None of the scenarios meet fullspace emissions limits.
The primary driver of emissions is investment in renewable generation and energy storage. Once these technologies are purchased, their use incurs no additional operational costs in the unit commitment model. However, the PCM is myopic in terms of emissions, and may take actions which lead to lower costs but higher emissions, such as export generation from thermal units in CAISO. As the fullspace model is run in discontinuous segments for reasons of computational tractability, it is impossible to effectively enforce emissions limits. It is difficult to say which, if any, of these fleets would be able to satisfy the emissions limits. Still, it is notable that 3-day representative periods present the lowest emissions, and longer, and thus fewer, periods have substantially higher emissions. 
This suggests that by modeling an intermediate-length period, interday energy storage can be leveraged to lower emissions. 
However, as the length of period grows, the number of periods must shrink. Without a sufficient number of periods, it is difficult to select periods which represent the annual behavior sufficiently well. This is illustrated by the higher emissions in the 4 and 5-day cases, which sample 9 and 7 periods respectively.



\begin{figure}[h]
\centering
\includegraphics[width = 0.95\linewidth]{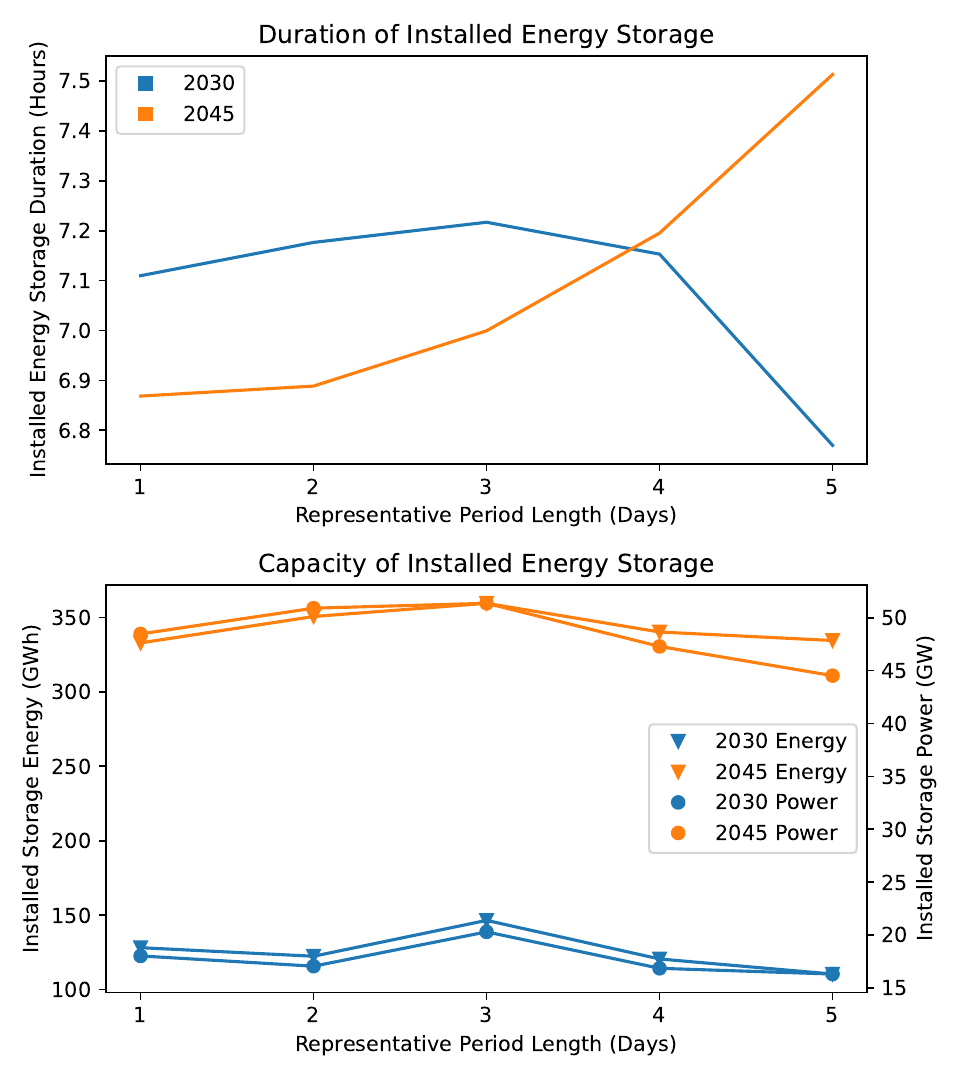}
\caption{Impact of representative period length on duration and capacity of installed energy storage.}
\label{fig:comboenergy}
\end{figure}

The duration of installed storage, and installed power and energy capacity of storage as a function of representative period length for 2030 and 2045 are shown in Fig. \ref{fig:comboenergy}. With regards to storage duration, the key takeaway is that increasing the length of representative period allows for utilization of storage for interday energy sharing, and the duration increases for lengths between 1 and 3 days. However, the tradeoff between number and length of representative seems to impact the ability of the surrogate days to effectively represent the full year, leading to less predictable effects with lengths over 3 days. With regards to the power and energy capacity of storage, a similar pattern is evident. Between lengths of 1 and 3 days, the installed capacities generally increase, and then begin to decrease again. This result is in line with the emissions result.

\begin{figure}[h]
\centering
\includegraphics[width = 0.95\linewidth]{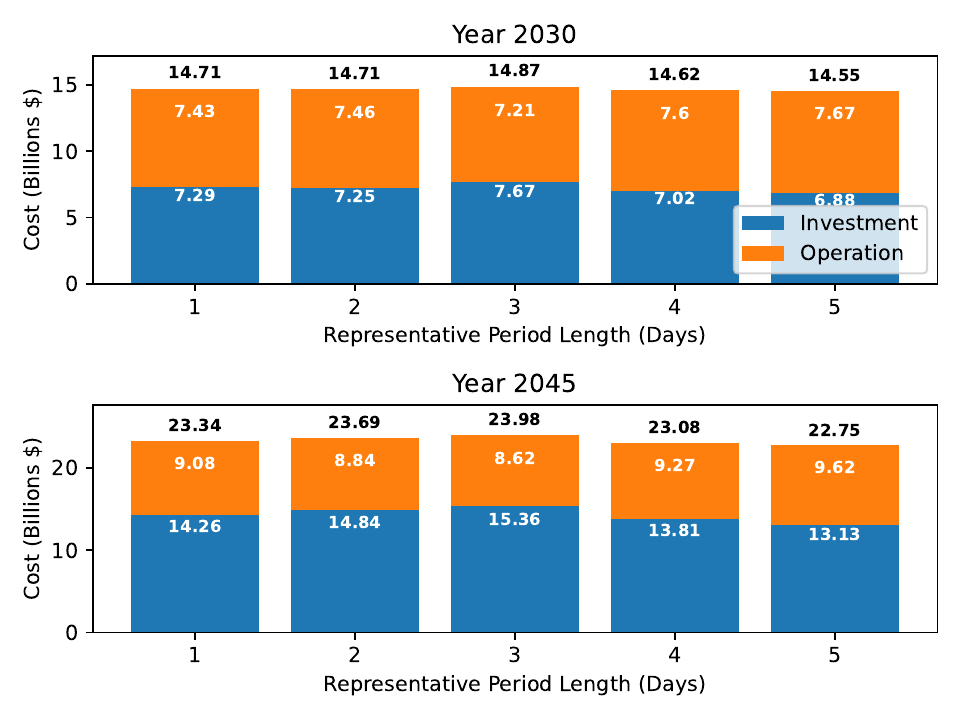}
\caption{Investment and operation costs of differing representative periods}
\label{fig:cost}
\end{figure}

Fig. \ref{fig:cost} shows the cost by year for each scenario. As one would expect from the emissions violations visualization, $d=4$ and $d=5$ have the lowest overall cost due to less build of renewable technologies. Most notable is that the $d=3$ result is very close to the $d=1$ result despite larger investment. This suggests that by representing longer periods, it is better able to capture the fullspace value of interday energy sharing. Thus, the cost of additional investment is offset by lower operating costs. Specifically, the total costs for 2030 are 1.1\% higher but have 7.1\% lower emissions. 

\begin{figure}[h]
\centering
\includegraphics[width = 0.95\linewidth]{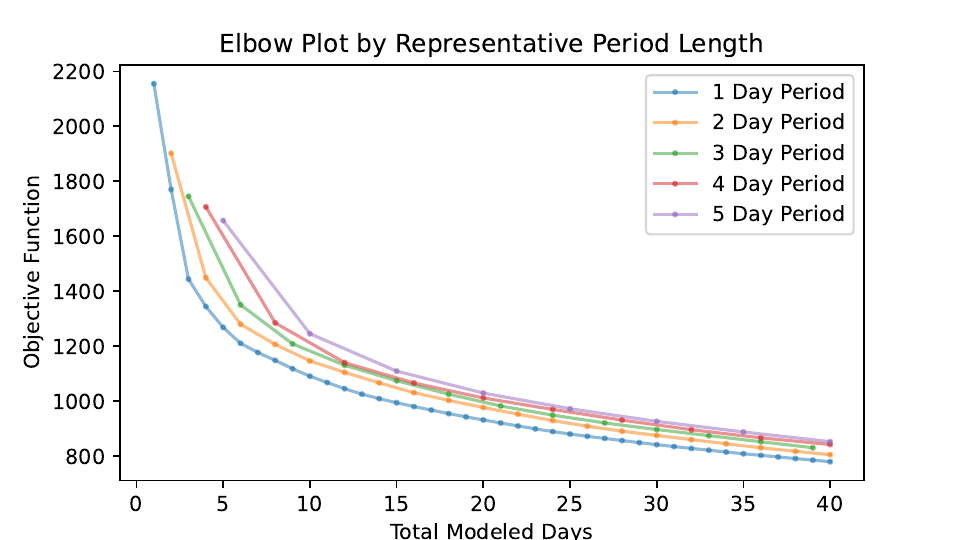}
\caption{Elbow plot of objective function \eqref{eq:algo} at different representative period lengths.}
\label{fig:elbow}
\end{figure}

Fig. \ref{fig:elbow} shows an elbow plot of the objective function of \eqref{eq:algo}. Intuitively, for a given number of total modeled days $k \times s / u$, the objective is best for more, shorter representative periods. The gap between the lines is larger at the lower total modeled days and begins to converge at higher. This characteristic explains why, for a fixed total modeled days, the representation degrades with higher period length. 

\section{Conclusion} \label{conclusion}
In this paper, we proposed a novel algorithm for selecting representative periods. The algorithm is particularly directed towards selecting periods longer than a single day, and is well suited for planning in systems with high penetration of variable renewable energy and reliance on energy storage. The method chooses representative days which minimize a distance measure to the timeseries of the full year. The proposed method was validated on a CEM based on California's decarbonization targets. The proposed method is competitive with the state-of-the-art for representative day selection, and we demonstrate the impact of representative period length on investment strategy.

\bibliographystyle{ieeetr}
\bibliography{references}

\end{document}